\documentclass[aps,twocolumn,showpacs,preprintnumbers,amssymb]{revtex4}

\newcommand{\be}{\begin{eqnarray}}
\newcommand{\ee}{\end{eqnarray}}

\newcommand{\eins}{\mbox{$1 \hspace{-1.0mm}  {\bf l}$}}
\def\bea{\begin{eqnarray}}
\def\eea{\end{eqnarray}}
\def\C{\hbox{$\mit I$\kern-.7em$\mit C$}}

\def\l{\langle}
\def\r{\rangle}

\def\tr{{\rm tr}}

\usepackage{dcolumn} 
\usepackage{bm} 

\begin{document}

\title{On the effective size of certain ``Schr\"{o}dinger cat'' like states}

\author{Wolfgang D\"{u}r$^1$, Christoph Simon$^2$ and J. Ignacio Cirac$^3$}

\affiliation{
$^1$ Sektion Physik, Ludwig-Maximilians-Universit\"at M\"unchen, Theresienstr.\ 37, D-80333 M\"unchen, Germany.\\
$^2$Centre for Quantum Computation, University of Oxford, Parks Road, Oxford OX1 3PU, U.K. \\ 
$^3$ Max--Planck Institut f\"ur Quantenoptik, Hans--Kopfermann Str. 1, D-85748 Garching,Germany.}

\date{\today}

\begin{abstract}
Several experiments and experimental proposals for the production of 
macroscopic superpositions naturally lead to states of the general form 
$|\phi_1\rangle^{\otimes N}+|\phi_2\rangle^{\otimes N}$, where the number of 
subsystems $N$ is very large, but the states of the individual subsystems have 
large overlap, $|\l \phi_1|\phi_2 \r|^2=1-\epsilon^2$. We propose two different 
methods for assigning an effective particle number to such states, using ideal 
Greenberger--Horne--Zeilinger (GHZ)-- states of the form $|0\r^{\otimes 
n}+|1\r^{\otimes n}$ as a standard of comparison. The two methods are based on 
decoherence and on a distillation protocol respectively. Both lead to an 
effective size $n$ of the order of $N \epsilon^2$.
\end{abstract}

\pacs{03.65.Ud, 03.67.-a, 03.65.Yz}

\maketitle
 


It was pointed out already in 1935 by Schr\"{o}dinger \cite{schroedinger} that 
quantum mechanics predicts the existence of superpositions of macroscopically 
distinct states. The observation of the related quantum interference effects is 
very difficult because of environment-induced decoherence. Nevertheless several 
methods for producing and verifying macroscopic superpositions have been 
proposed, in systems ranging from superconductors \cite{leggett} over 
Bose-Einstein condensates (BECs) \cite{cirac,ruostekoski} and opto-mechanical 
systems \cite{bose} to small cantilevers coupled to superconducting islands 
\cite{armour}. Recently there have even been the first experimental 
demonstrations of the superposition of distinct macroscopic current states in 
superconducting quantum interference devices (SQUIDs) \cite{squidexp}. 

The states produced in such proposed experiments can often be described to a good approximation by 
\be
|\psi\r=\frac{1}{\sqrt{K}}(|\phi_1\rangle^{\otimes N}+|\phi_2\rangle^{\otimes N}),
\label{epsiloncat}
\ee
with $K=2+\l \phi_1|\phi_2 \r^N + \l \phi_2|\phi_1\r^N$. Here $|\psi\r$ is a 
state of $N$ two-level systems (qubits). The individual qubits could be seen as 
simple models for many different physical systems, including the atoms in a BEC 
inside a double-well potential \cite{cirac}, atoms in two internal states
or
the Cooper pairs 
in a SQUID (which can flow in clockwise or anti-clockwise direction).

The essential point for our present discussion is that the states $|\phi_1\r$ 
and $|\phi_2\r$ are not necessarily orthogonal. In fact, we will study the case 
where $|\l \phi_1|\phi_2 \r|^2=1-\epsilon^2$ is very close to one ($\epsilon$ is 
small). Note that in spite of this the overlap between the two terms in 
(\ref{epsiloncat}) can be very small for large $N$, since it is given by $|\l 
\phi_1|\phi_2 \r|^{2N}=(1-\epsilon^2)^N$, which is well approximated by 
$e^{-N\epsilon^2}$ for small $\epsilon$.

We investigate
 how states of the form (\ref{epsiloncat}) compare to ideal GHZ states of the form
\be
|GHZ_n\rangle = \frac{1}{\sqrt{2}}(|0\r^{\otimes n}+|1\r^{\otimes n}).
\label{ghz}
\ee
States of the form (\ref{ghz}) can be produced to good approximation in quantum 
optical systems, including atoms in cavity quantum electrodynamics \cite{haroche},  
trapped ions \cite{wineland} and photons from parametric down-conversion 
\cite{zeilinger}. So far, particle numbers $n$ up to 4 have been 
achieved. On the other hand, states of the form (\ref{epsiloncat}) can involve 
macroscopic numbers of particles, in combination with small values of 
$\epsilon$. An important question is whether there is a well-defined way to 
compare these two -- very different -- cases. That is, can one give a meaningful 
answer to the question whether e.g. a state of the form (\ref{epsiloncat}) with 
$N=10^6$ particles, but $\epsilon=10^{-3}$ is more or less entangled than an 
ideal GHZ state with $n=10$?

A first simple way of assessing the ``size'' of states of the form 
(\ref{epsiloncat}) is to look at the overlap between the two terms. However, as 
pointed out above, this will be close to zero in most interesting cases and thus 
does not lead to a very sensitive criterion. A very intuitive way of comparing 
(\ref{epsiloncat}) and (\ref{ghz}) would be to assign to (\ref{epsiloncat}) an 
``effective particle number'' $n$, which could be interpreted as saying that the 
state (\ref{epsiloncat}) is (in a certain well-defined sense) {\it equivalent} 
to an ideal GHZ state of $n$ qubits. This requires well-defined and physically 
meaningful methods of determining such an effective particle number. Here we 
will propose two such methods and show that they lead to essentially the same 
result, namely that the effective $n$ of a state of the form (\ref{epsiloncat}) 
is of the order of $N \epsilon^2$.

Our two methods are very different. The first one is based on the rate of 
decoherence. An important potential application of states of the form 
(\ref{epsiloncat}) is for the observation of (or the search for) weak 
decoherence processes, ultimately including unconventional ones as predicted by 
spontaneous wave-function collapse models \cite{collapse}. If the state 
(\ref{epsiloncat}) is as sensitive to decoherence as an ideal $n$-qubit GHZ 
state, 
it is natural to say that its effective size is $n$.  

The second method of assigning an effective size to our states is in the spirit 
of quantum information, viewing multi-party entanglement as a convertible 
resource. We ask how much ideal GHZ entanglement can be {\it distilled} from the 
states (\ref{epsiloncat}) by local operations (acting separately on each qubit) 
and classical communication. 

It is remarkable that these two very different approaches lead to the same 
result. We believe that this suggests that $N \epsilon^2$ is indeed a good 
physical quantification of the ``size'' of macroscopic superpositions of the 
present type. We conclude by giving another simple argument in favor of the 
proposed scaling of the effective size, based on particle loss.


Let us now follow our first approach and study the effect of local decoherence 
on the state (\ref{epsiloncat}). We will consider the case where each of the 
particles undergoes an independent decoherence process, i.e. each particle is 
coupled to an independent  bath. The effect of decoherence is quantified as the 
rate of decay of the off-diagonal elements in the natural basis. Note that we 
are interested in properties of states and not of physical set--ups which 
generate them. In this sense, although for different physical systems the 
decoherence process may be completely different, we can study the behavior of 
the states describing those systems under a certain decoherence process in order 
to compare the properties of these states.

We consider phase decoherence in the natural basis, that is diagonal elements in 
the $\{|0\rangle,|1\rangle\}$ basis remain unchanged, while off diagonal 
elements $|0\rangle\langle1|, |1\rangle\langle0|$ decay with a rate $\gamma$. 
The decoherence process of an individual system is described by $|i\rangle\langle 
j| \rightarrow e^{-\gamma t} |i\rangle\langle j|$ for $i\not= j$ and 
$|j\rangle\langle j| \rightarrow |j\rangle\langle j|$, which corresponds ---in a 
quantum information language--- to a dephasing channel. The action of this 
channel may be described by the completely positive map ${\cal E}$ defined 
through ${\cal E}(\rho) = p_0  \rho + (1-p_0) \sigma_z \rho \sigma_z$, where 
$p_0=(1+e^{-\gamma t})/2$ and $\sigma_z$ is a Pauli matrix.  

It is straightforward to establish the effect of this decoherence process on an 
ideal $n$--particle GHZ-state. The density matrix for the state Eq. (\ref{ghz}) 
is $1/2(|0\r\l 0|^{\otimes n}+|1\r\l 1|^{\otimes n}+|0\r\l 1|^{\otimes n}+|1\r\l 
0|^{\otimes n})$. Since ${\cal E}^{\otimes n}(\sigma ^{\otimes n}) =({\cal 
E}(\sigma))^{\otimes n}$, one  can study the decay of the  off-diagonal elements 
$|0\rangle\langle 1|^{\otimes n}$ and $|1\rangle\langle 0|^{\otimes n}$ by 
considering the action of the decoherence channel ${\cal E}$ on the single-qubit 
operators $|0\r\l 1|$ and $|1\r\l 0|$. To be specific, we consider the trace 
norm, $||A||_1\equiv \tr \sqrt{A^\dagger A}$, of $a_t={\cal E}(a_0)$, where 
$a_0=|0\rangle\langle 1|$.  Note that $||A^{\otimes N}||_1 = ||A||_1^N$. Since 
$a_t= e^{-\gamma t} a_0$, we have that $|| a_t^{\otimes n} ||_1 = 
||a_t||_1^n=e^{-\gamma n t} ||a_0||_1$. The off-diagonal element of the GHZ 
state decays with a rate $\gamma n$.

We want to compare this to the decay rate of the off-diagonal terms for the 
$N$--particle states $|\psi\rangle$ of the form (\ref{epsiloncat}). Without loss 
of generality we denote 
\bea
|\phi_1\rangle &=&|0\rangle, \nonumber \\
|\phi_2\rangle &=&\cos(\epsilon)|0\rangle + \sin(\epsilon) |1\rangle, \label{epsiloncat2}
\eea
and use the shorthand notation $c_\epsilon\equiv \cos(\epsilon)$ and 
$s_\epsilon\equiv \sin(\epsilon)$. We have that $K=2(1+c_\epsilon^N)$ and for 
small $\epsilon$, $|\l \phi_1|\phi_2 \r|^2=c_\epsilon^2 \approx 1-\epsilon^2$. 
The density matrix for the state Eq. (\ref{epsiloncat}) is 
$1/K(|\phi_1\r\l\phi_1|^{\otimes N}+|\phi_2\r\l\phi_2|^{\otimes 
N}+|\phi_1\r\l\phi_2|^{\otimes N}+|\phi_2\r\l\phi_1|^{\otimes N})$. 
We are interested in the decay rate of the 
off--diagonal elements. As before, the problem can be reduced to studying 
single-qubit operators, namely $b_0\equiv |\phi_1\r \langle \phi_2|=c_\epsilon 
|0\r\l 0| + s_\epsilon |0\r \langle 1|$. We have that $b_0$ changes due to the 
above decoherence process to $b_t={\cal E}(b_0)=\sqrt{d} |0\r \langle \chi_t|$ 
with $ |\chi_t \r = 1/\sqrt{d} (c_\epsilon |0\rangle + s_\epsilon e^{-\gamma t} 
|1\r )$ and $d=c_\epsilon^2 +s_\epsilon^2 e^{-2 \gamma t}$, such that 
$|\chi_t\r$ is properly normalized. It follows that $||b_t^{\otimes N}||_1 = 
d^{N/2}$. For $\epsilon \ll 1, t \ll \gamma^{-1}$, $N\epsilon^2 \gamma t \ll1$, we have $d\approx 
1-2\epsilon^2\gamma t$ and thus $d^{N/2}\approx e^{-\gamma N \epsilon^2 t}$. 
This implies that the rate with which the coherences of the state $|\psi\rangle$ 
decay is given by $\gamma N \epsilon^2$. That is, the decoherence rate of a 
state of the form Eq. (\ref{epsiloncat}) with $|\langle \phi_1|\phi_2\rangle|^2 
= 1-\epsilon^2$ is the same as that of an ideal $n$--party GHZ state with $n=N 
\epsilon^2$ and thus one may associate an effective particle number $n=N 
\epsilon^2$ to the state $|\psi\rangle$.  

The observed decoherence rate is not restricted to this specific 
decoherence model. 
Consider for example the basis independent decoherence model of a partially 
depolarizing channel.  In this case, the individual decoherence process for each 
qubit is described by $|i\rangle\langle j| \rightarrow \mu |i\rangle\langle j| + 
(1-\mu) \delta_{i,j} \frac{1}{2}\eins$ where $\mu \equiv e^{-\gamma t}$. 
Equivalently, the completely positive map $\tilde{\cal E}$ describing this 
process is given by $\tilde{\cal E}(\rho) = \sum_{i=0}^3 p_i \sigma_i \rho 
\sigma_i$ with $p_0=(3\mu+1)/4$ and $p_1=p_2=p_3=(1-\mu)/4$, where 
$\sigma_0=\eins$, and the $\sigma_i$ are Pauli matrices. We find that $a_0$ [$b_0$] 
changes due to this decoherence process to $a_t=\tilde{\cal E}(a_0)= \mu a_0$ 
[$b_t =\tilde{\cal E}(b_0)= c_\epsilon (1+\mu)/2 |0\r \langle 0| + (1-\mu)/2  
|1\r \langle 1| + s_\epsilon \mu  |0\r \langle 1|$]. 
One obtains that 
$||a_t||_1= \mu = e^{-\gamma t}$ and 
$||b_t||_1=\sqrt{c_\epsilon^2+\mu^2 s_\epsilon^2}=\sqrt{d}$, which is exactly the same as 
in the case of the dephasing channel. One thus recovers exactly the same decoherence 
rates --$\gamma n$ and $\gamma N \epsilon^2$ respectively-- as in the case of 
the dephasing channel.


Let us now turn to our second approach, which is more in the spirit of quantum 
information. We again consider states $|\psi\rangle$ of the form 
(\ref{epsiloncat}) with $|\phi_{1,2}\rangle$ defined in Eq. (\ref{epsiloncat2}). 
We are interested in the {\it distillation} of ideal $n$--particle GHZ 
states (\ref{ghz}) from these states under the condition that only local 
operations and classical communication are allowed.  The restriction to local 
operations is essential if one wants to quantify the entanglement contained in a 
given state because non-local operations could create additional entanglement. 
We are only interested in the number of particles which form a GHZ state after 
the distillation process, i.e. the effective size of the GHZ-state, and 
not which of the particles are entangled. 

We show that the average number of the particles which is in an ideal 
GHZ--state after the distillation process scales essentially like $n=N \epsilon^2$. We  
(i) provide an explicit protocol to produce ---with unit probability--- $n$--party GHZ states from a single 
copy of $|\psi\rangle$ where the average value of $n$ is $N \epsilon^2/2$ and (ii) 
show that even in the asymptotic limit, i.e. considering several identical 
copies of the state $|\psi\rangle$ , this average value 
is bounded from 
above by $n\approx N \epsilon^2 (-\log_2(\epsilon)/2)$ \cite{noteLog}.


Let us start with (i), a practical protocol which transforms a single copy of 
$|\psi\rangle$ deterministically into $n$--party GHZ states by means of local 
filtering measurements. The protocol we propose is a generalization to 
multipartite systems of the distillation protocol of Ref. \cite{Ac00} for the 
optimal distillation of tripartite GHZ states from a single copy of an arbitrary 
pure state of three qubits. Consider the local filtering measurement described 
by the operator $A\equiv 
k(|0\rangle\langle\tilde{\phi_1}|+|1\rangle\langle\tilde{\phi_2}|)$, where 
$\{|\tilde{\phi_1}\r,|\tilde{\phi_2}\rangle\}$ is the biorthonormal basis to 
$\{|\phi_1\rangle,|\phi_2\rangle\}$, i.e. $\langle \phi_j|\tilde{\phi_l}\rangle= 
\delta_{jl}$. The constant $k$ is chosen such that the other operator $\bar A$ 
of the local, two--outcome generalized measurement $\{A,\bar A\}$ ---which fulfills $A^\dagger A +\bar 
A^\dagger \bar A =\eins$--- has rank one. This implies that in case one obtains 
the outcome corresponding to $A$, then $A\otimes \eins^{\otimes 
N-1}|\psi\rangle= k/\sqrt{K} (|0\rangle|\phi_1\rangle^{\otimes 
N-1}+|1\rangle|\phi_2\rangle^{\otimes N-1})$, while for the other outcome $\bar A\otimes 
\eins^{\otimes N-1}|\psi\rangle \propto |\chi\rangle \otimes 
(|\phi_1\rangle^{\otimes N-1}+|\phi_2\rangle^{\otimes N-1})$, i.e. the measured 
particle factors out. The distillation protocol works as follows: Each of the 
parties performs locally the two--outcome generalized measurement $\{A,\bar A\}$ 
and all those $n$ parties which obtained a positive outcome, i.e. the outcome 
corresponding to $A$, finally share an ideal $n$--party GHZ--state of the form 
(\ref{ghz}), while the remaining parties are in a product state. We shall be 
interested in the expectation value of $n$. 

Using the notation of 
Eqs. (\ref{epsiloncat}, {\ref{epsiloncat2}) 
, we find that 
\be
A=\frac{\sqrt{1-c_\epsilon}}{s_\epsilon} \left( \begin{array}{cc} 
s_\epsilon & -c_\epsilon \\ 0 & 1 \end{array}  \right).
\ee
In the $j^{\rm th}$ measurement, the probability to obtain the outcome 
corresponding to $A$ is given by $p=(1-c_\epsilon)/(1+c_\epsilon^{N-j+1})$ 
provided that none of the previous measurements was successful. In case one of 
the previous measurements was already successful, the probability to obtain an 
outcome corresponding to $A$ is given by $\tilde p = (1-c_\epsilon)$ for the 
remaining parties. This different behavior after the first successful 
measurement can be easily understood by noting that once one of the measurements 
was successful, then the (normalized) state after the measurement is given by 
$1/\sqrt{2}(|0\rangle|\phi_1\rangle^{\otimes 
N'}+|1\rangle|\phi_2\rangle^{\otimes N'})$, while otherwise the normalization 
constant is $K=2(1+c_\epsilon^{N-j+1})$. One finds that the probability $q_n$ to 
obtain $n$ (where $n\geq 1$) successful measurements ---and thus $n$--party GHZ 
states--- is given by 
\be
q_n= (1-c_\epsilon)^n c_\epsilon^{N-n} \left( \begin{array}{c} N \\ n \end{array} \right) \frac{1}{1+c_\epsilon^N},
\ee
while $q_0=2c_\epsilon^N/(1+c_\epsilon^N)$. Note that this probability distribution is ---up 
to the factor $1/(1+c_\epsilon^N)$ and correspondingly the value of $q_0$ --- very similar to a binomial distribution. The 
expectation value $<n> \equiv \sum_{j=0}^{N} q_j j$ is given by $<n>= 
(1-c_\epsilon) N/(1+c_\epsilon^N)$ which simplifies for $\epsilon \ll 1$ and 
$N\epsilon^2 \gg 1$ to
\be
<n> \approx N\epsilon^2 /2.
\ee
This provides the desired lower bound for the distillation rates of $n$--party GHZ--states.


Regarding (ii), the announced upper bound for the distillation rate, we use the 
fact that the von--Neumann entropy of the reduced density operator with respect 
to system 1 $\rho_1$ \cite{notePT}, $S_1(\rho_1)\equiv -\tr(\rho_1 \log_2 
\rho_1)$, is an entanglement monotone, i.e. not increasing under local 
operations and classical communication \cite{Be96,Be99,Vi00J}.  

We consider the distillation process in the asymptotic limit, i.e. the 
transformation of $M\rightarrow \infty$ identical copies of the state $|\psi\rangle$ to 
$n$--particle GHZ states \cite{Be99}. Such a distillation protocol consists of an arbitrary 
sequence of local operations (including measurements), possibly assisted by 
classical communication, mathematically described by a multi--local superoperator 
\cite{Be99}. The protocol produces a certain number, say $M_n$ copies, of 
$n$--party GHZ--states, which can ---as we are only interested in the number of 
parties which constitute a GHZ state--- without loss of generality be considered 
to be symmetrically distributed among the $N$ parties \cite{noteProtocol}. 
Such a symmetric configuration 
is denoted by $|\overline{GHZ}_n\r^{\otimes M_n}$ \cite{noteGHZ}. The 
distillation protocol is described by the following transformation
\be
|\psi\rangle^{\otimes M} \rightarrow \bigotimes_{n=2}^{N} |\overline{GHZ}_n\rangle^{\otimes p_n M},
\ee
where $p_n\geq 0$ denotes the average number of $n$--party GHZ states which are produced per copy 
from $|\psi\rangle$. 

Given  the monotonicity of the entropy under local operations, we obtain
\be
M S_1(|\psi\rangle) \geq \sum_{n=2}^{N} S_1(|\overline{GHZ}_n\r^{\otimes p_n M}),
\ee
where $S_1(|\overline{GHZ}_n\rangle^{\otimes p_n M})$ denotes the entropy of the 
reduced density operator with respect to system 1 of $p_n M$ copies of 
(symmetrically distributed) $n$--particle GHZ states \cite{noteGHZ}. Since the 
probability that the first particle belongs to the $n$ entangled particles is 
given by $p=\left(\begin{array}{c} N-1 \\ n-1 \end{array}\right) / 
\left(\begin{array}{c} N \\ n \end{array}\right)= n/N$, we have that 
$S_1(|\overline{GHZ}_n\rangle^{\otimes p_n M}) = p_n M n/N$ and thus 
$\sum_{n=2}^{N} p_n n \leq N S_1(|\psi\rangle)$. It is straightforward to 
calculate $S_1(|\psi\rangle)$ using that for $|\psi\rangle$ (\ref{epsiloncat}), 
the reduced density operator with respect to system 1 is given by $\rho_1= 
[(1+c_\epsilon^2+2c_\epsilon^N) |0\r\langle 0| +  s_\epsilon c_\epsilon 
(1+c_\epsilon^{N-2}) (|0\r\langle 1| +  |1\r\langle 0|) + s_\epsilon^2 
|1\r\langle 1|] 1/(2+2c_\epsilon^N)$.  For $\epsilon \ll 1$ and $N \epsilon^2 
\gg 1$, one obtains that $S_1(|\psi\r) \approx -\epsilon^2 \log_2(\epsilon)/2$ 
which implies
\be
\sum_{n=2}^{N} p_n n \leq -N \epsilon^2 \frac{1}{2} \log_2(\epsilon)
\ee
as announced \cite{noteDist}.

We would finally like to mention another simple argument that suggests the same 
scaling for the effective size of the states Eq. (\ref{epsiloncat}). Let us 
compare the effects of particle loss on the state (\ref{epsiloncat}) and on an 
ideal GHZ-state. Suppose that every qubit is lost with a probability $\lambda$. 
Consider an $n$-qubit GHZ state. As soon as a single qubit is lost, the state 
becomes completely diagonal. There is only an off-diagonal element in the case 
of no losses, which has a probability of $(1-\lambda)^n$. The expectation value 
of the off-diagonal element in the case of losses is therefore 
$\frac{1}{2}(1-\lambda)^n$, equal to $\frac{1}{2}e^{-\lambda n}$ for small 
$\lambda$.

On the other hand, for the state (\ref{epsiloncat}), tracing out particles 
reduces the size of the off-diagonal terms but does not completely remove them. 
Tracing out $k$ particles multiplies the off-diagonal terms by a factor of $\l 
\phi_1|\phi_2\r^k=(1-\epsilon^2/2)^k$, equal to $e^{-k \epsilon^2/2}$ for small 
$\epsilon$. The typical number of particles lost will be $N \lambda$, therefore 
the typical off-diagonal term will go like $e^{-\lambda N \epsilon^2/2}$. For 
large $N \lambda$ the probability distribution will be strongly peaked around 
the typical value. Therefore the expectation value of the off-diagonal term will 
be of order $e^{-\lambda N \epsilon^2/2}$. One sees that if the ideal GHZ state 
has $n=N \epsilon^2/2$, then the expectation values of the off-diagonal terms 
will have the same size for the two states. This is one more confirmation for 
our proposal that the ``effective size'' of the state (\ref{epsiloncat}) scales 
like $N \epsilon^2$. Note that  there are several other arguments which confirm 
this effective size $n\approx N\epsilon^2$, e.g. an argument related to the 
statistical distinguishability of states $|\phi_{1,2}\r$ as pointed out in Ref. 
\cite{Ac02}.


To summarize, we provided two different methods to assign an effective particle 
number to GHZ--like states of the form $|\psi\rangle \propto 
|\phi_1\rangle^{\otimes N}+|\phi_2\rangle^{\otimes N}$ with 
$|\langle\phi_1|\phi_2\r|^2=1-\epsilon^2$. The first method is based on the rate 
of decoherence, and we found that $|\psi\r$ behaves like an ideal $n$--party GHZ 
state with $n\approx N \epsilon^2$. In the second method, which is more in the 
spirit of quantum information, we provided lower and upper bounds for the 
distillation rates $p_n$ of ideal $n$--party GHZ--states using only local 
operations and classical communication. Again, we found that $\sum p_n n \approx 
N \epsilon^2$, i.e. the average number of particles which form an ideal 
$n$--party GHZ--state essentially scales like $n\approx N \epsilon^2$. This 
illustrates that not only the number of particles but also the properties of the 
states appearing in the microscopic description of the system determine the 
effective size of the corresponding Schr\"odinger cat like state.

Some open questions remain. On the one hand, we have considered a particular 
class of Schr\"odinger cat like states (\ref{epsiloncat}). One can have physical 
systems where the macroscopic superposition cannot be described by states of 
this form. In particular, the states appearing in the superposition may not be a 
tensor product of identical microscopic states, $|\phi_{1,2}\r^{\otimes N}$, 
either because they are entangled themselves ---e.g. the position states of the 
atoms in an oscillating micro-mechanical cantilever or mirror--- or the the 
corresponding system cannot be decomposed in a natural way into subsystems 
---e.g. a superposition of two coherent states, $|\alpha\r+|-\alpha\r$. It would 
be interesting to study by similar means the degree of ''catness'' for those 
systems and maybe compare them with the ones treated here.

On the other hand, there are experiments where macroscopic superpositions have 
been created but there is no microscopic description of the states produced 
\cite{squidexp}. It would be interesting to find such a microscopic description 
and ---in case the states can be written as (\ref{epsiloncat}) (which would be 
natural)--- to assess an effective size of the states for these experiments.


We would like to thank A. Acin for useful discussions.
This work was supported by European Union under project EQUIP (contract 
IST-1999-11053) and through the Marie Curie fellowships HPMF-CT-2001-01209 (W.D.)
and HPMF-CT-2001-01205 (C.S.),  the ESF, and the Institute for Quantum Information GmbH.





\begin{thebibliography}{99}
\bibitem{noteLog}
If $\epsilon$ 
is 
e.g. between $10^{-15}$ and $10^{-2}$, this upper bound ranges between
$25N \epsilon^2$ and $3N\epsilon^2$. For $\epsilon=2^{-10}\approx10^{-3}$, we have 
$-\log_2(\epsilon)/2=5$, while $\epsilon^2 \approx 10^{-6}$.
\bibitem{notePT}
The reduced density operator $\rho_1$ with respect to system 1 of an 
$N$--partite pure state $|\Psi\rangle_{12\ldots N}$ is defined as $\rho_1\equiv 
\tr_{2,3,\ldots N} (|\Psi\r \langle \Psi|)$.
\bibitem{noteProtocol}
By forming $l$ groups of $M/l$ copies of the (symmetric) state $|\psi\r$ and 
choosing in each group one of the $l=N!$ possible labelings of the parties 
before applying the given (possibly asymmetric) distillation protocol to each 
group independently, one achieves that the new protocol obtained in this way 
produces $n$--particle GHZ states with the same rate as the initial protocol, 
and the distribution of the $n$--particle GHZ states among the $N$ parties is 
symmetric $\forall n$.
\bibitem{noteDist}
One can use the above method to 
obtain also a bound for each $p_n$, i.e. assuming that only $n$--party GHZ 
states for some fixed $n$ are created by the distillation protocol. One obtains $p_n \leq -N 
\epsilon^2  \log_2(\epsilon) /(2n)$ and finds that $p_n \approx 1$ if $n\leq -N 
\epsilon^2  \log_2(\epsilon)/2$, i.e. only in this case it is in principle possible to produce $n$--party 
GHZ states with a rate per copy 
approximately equal to 1.
\end{thebibliography}
\end{document}